\pdfoutput=1 
\documentclass[aps,twocolumn,preprintnumbers,floatfix,nofootinbib]{revtex4}

\usepackage{graphicx}
\usepackage{dcolumn}
\usepackage{bm}
\usepackage{epsfig}
\usepackage{color}
\usepackage{multirow}

\usepackage{amsmath}
\usepackage{amssymb}

\def\fun#1#2{\lower3.6pt\vbox{\baselineskip0pt\lineskip.9pt
  \ialign{$\mathsurround=0pt#1\hfil##\hfil$\crcr#2\crcr\sim\crcr}}}

\input epsf

\newcommand{\be}{\begin{equation}}
\newcommand{\ee}{\end{equation}}
\newcommand{\bea}{\begin{eqnarray}}
\newcommand{\eea}{\end{eqnarray}}

\begin{document}


\title{Looking for a light Higgs boson in the overlooked channel}


\author{
James S. Gainer$^{a,b}$, Wai-Yee Keung$^{c}$,  Ian Low$^{a,b}$, 
and Pedro Schwaller$^{a,c}$ }
\affiliation{
\vspace*{.1cm}
$^a$ \mbox{High Energy Physics Division, Argonne National Laboratory, 
Argonne, IL 60439}\\
$^b$ \mbox{Department of Physics and Astronomy, Northwestern University, 
Evanston, IL 60208} \\
$^c$  \mbox{Department of Physics, University of Illinois at  Chicago, IL 
60607}\\
}

\begin{abstract}
\vspace*{-0.6cm}
The final state obtained when a Higgs boson decays to a photon and a $Z$ boson has 
been mostly overlooked in current searches for a light Higgs boson. 
However, when the $Z$ boson decays leptonically, all final state particles in this 
channel can be measured, allowing for accurate reconstructions 
of the Higgs mass and angular correlations.  We determine the sensitivity of the
Large Hadron Collider (LHC) running at center of masses energies of
$8$ and $14$ TeV  to Standard Model (SM) Higgs bosons with masses in the $120 - 130$ GeV range.
For the $8$ TeV LHC, sensitivity to several times the the SM cross section times branching ratio
may be obtained with $20$ inverse femtobarns of integrated luminosity, while for the $14$ TeV
LHC, the SM rate is probed with about $100$ inverse femtobarns of integrated luminosity.
\end{abstract}


\maketitle


{\bf Introduction} -- The search for the Higgs boson 
is entering a critical phase. Data collected at the LHC 
rules out the SM Higgs boson for a wide range of masses 
and may suggest a Higgs boson with mass near $125$ 
GeV~\cite{ATLAS:2012ae, Chatrchyan:2012tx}.  
Searches for a light SM Higgs in the still-relevant mass window rely primarily on 
the $\gamma\gamma$ channel, though the $WW^*\to 2\ell 2\nu$ channel and  
the golden channel, $ZZ^*\to 4\ell$, are also important. 

So far very little attention has been given to the 
$Z\gamma\to \ell\bar{\ell}\gamma$ channel~\cite{Cahn:1978nz}, although its event rate is 
comparable to that of the golden channel for a light SM Higgs boson. 
Nevertheless, this channel has the advantage that all final state particles 
can be measured well, which carries several important implications: 
1) the Higgs mass could be measured from the total invariant mass spectrum, 
2) the spin of a putative signal can be determined by studying angular
 correlations~\cite{arXiv:1010.2528}, and 3) the separation of signal from 
background can be facilitated by employing full kinematic information, potentially
allowing searches with enhanced sensitivities.
For the golden channel in $ZZ^*\to 4\ell$ the above questions have been 
studied extensively \cite{Matsuura:1991pj,Keung:2008ve, Gainer:2011xz}, 
but we are not aware of any detailed studies for the $Z\gamma$ channel.

Measurements of all four Higgs decay modes into electroweak bosons are 
in fact very important in determining the electroweak quantum numbers of 
a putative Higgs signal \cite{arXiv:1005.0872}.  Furthermore, an electroweak 
singlet scalar could easily have a branching fraction in the $Z\gamma$ mode 
that is orders of magnitude larger than the SM 
expectation~\cite{arXiv:1105.4587} which provides an important additional 
incentive for studying this channel. 

{\bf Kinematics} -- The kinematics of $Z\gamma\to \ell\bar{\ell}\gamma$ is 
described by three angles, $\Theta$, $\theta$ and $\phi$, where
$\Theta$ may be taken to be the 
angle describing the production of the $Z$ boson in the center of mass frame,
and $\theta$ and $\phi$ are the angles that describe the decay of the $Z$ to leptons,
as defined in more detail in~\cite{Gainer:2011xz}.  
To accomodate events with jet radiation in the final state, we use the 
momentum of the $Z\gamma$ system in the lab frame, 
rather than the beam axis, to define $\Theta$.  

%

\begin{figure}[t]
\includegraphics[scale=0.42, angle=0]{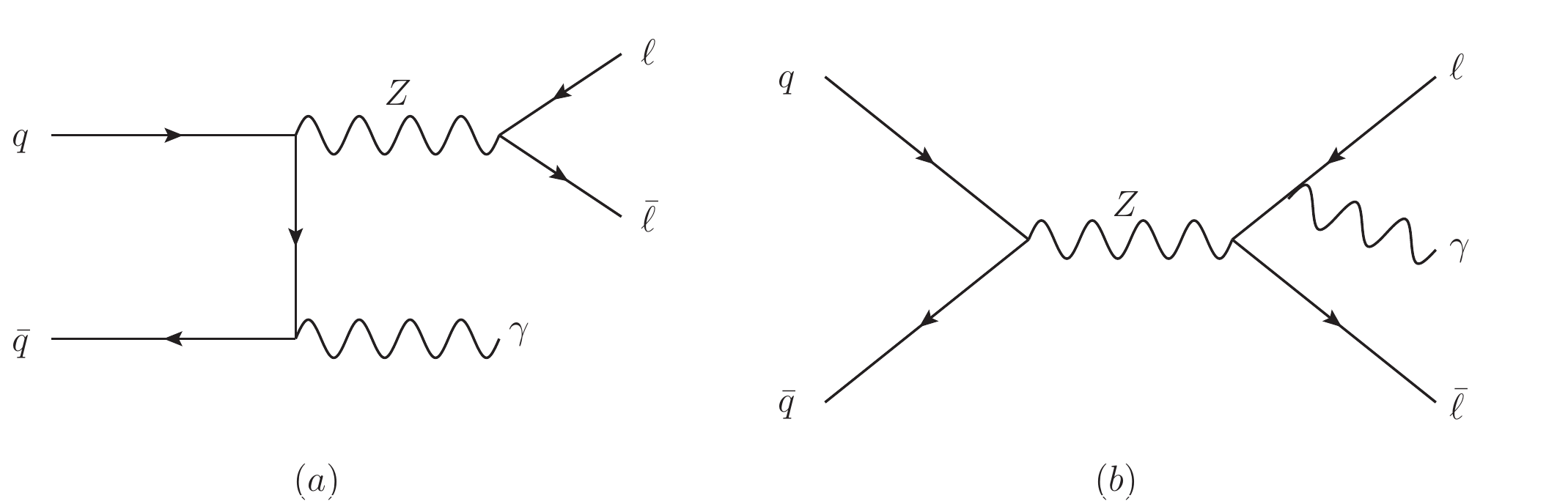}  
\caption{\label{fig1}{\em Feynman diagrams contributing to 
$q\bar{q}\to \ell\bar{\ell}\gamma$ are shown in (a) and (b).}}
\end{figure}

The dominant irreducible background to the Higgs signal arises from initial 
state radiation (ISR) and final state radiation (FSR) from Drell-Yan 
production of a $Z$ boson; the diagram describing this process 
is shown in Fig.~\ref{fig1} (a) and (b). The invariant mass of the
$Z\gamma$ system from FSR events is close to the $Z$ boson mass,
so this background is removed efficiently by imposing $m_{\ell\ell\gamma}>100$~GeV,
and we can focus on  the ISR diagram and the 
corresponding $u$-channel diagram for the rest of this analysis. 


The signal and background cross sections were computed using the helicity basis 
in~\cite{Hagiwara:1986vm}.
We now discuss some qualitative features of these
differential cross sections, in particular the $\Theta$ dependence of the signal
and background processes.
%

\begin{figure*}[ht]
\includegraphics[width=0.9\textwidth, angle=0]{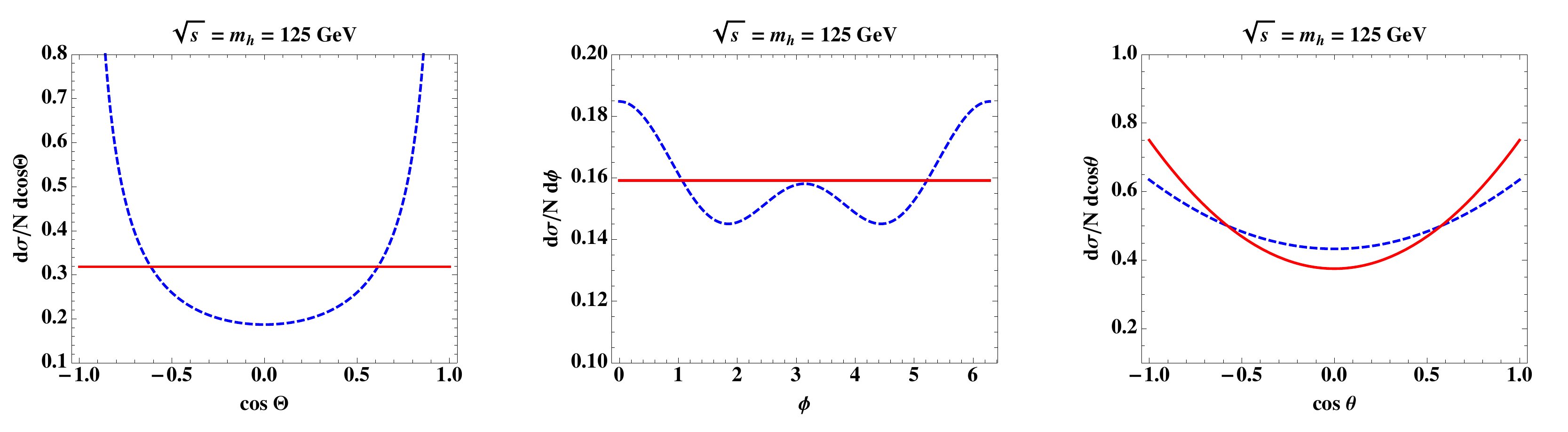}  
\caption{\label{fig:dist}{\em Signal (red, solid) and background (blue, dashed) distributions in $\cos \Theta$, $\phi$ and $\cos \theta$, with $\sqrt{\hat{s}} = m_h = 125$~GeV.}}
\end{figure*}

In the signal case, angular distributions follow from the fact that the Higgs is a scalar particle,  and hence
only the decay angle $\theta$ has a nontrivial distibution:
\be
\label{signal diff}
\frac1{N} \frac{d\sigma}{d\cos\Theta\, d\cos\theta\, d\phi} = (1+\cos^2\theta) \ ,
\ee
%
%
For the background distributions, the non-vanishing helicity 
combinations are $(\lambda_1,\lambda_2)=(\pm, \mp), (0,\pm)$, 
and  $(\pm, \pm)$.  The production angular distribution 
exhibits a collinear singularity at $\cos\Theta=\pm 1$, which is seen by 
examining the $t$-channel propagator in Fig.~\ref{fig1} (a),
\be
\frac{1}{(k_{\bar{q}}-p_\gamma)^2} = -\frac{1}{2E_{\bar{q}} E_\gamma 
(1-\cos\Theta)} \ ,
\ee
while the $u$-channel propagator gives the collinear singularity at 
$\cos\Theta=-1$. Thus the production angular distribution for the background 
process is peaked at $\cos\Theta=\pm 1$, producing forward and backward photons.  
The singularity 
is removed by the $p_T$ cuts on the photon and leptons. Explicit calculations lead to 
\bea
\label{bkdg diff}
&&\frac1{N'}\frac{d\sigma}{d\cos\Theta\, d\cos\theta\, d\phi}= \nonumber \\
&&  \ (g_r^2+g_\ell^2)(g_R^2+g_L^2) \, {\cal G}_1 + (g_r^2-g_\ell^2)(g_R^2-g_L^2) {\cal G}_2 \,, 
\eea
with
\bea
 {\cal G}_1 &= & \left[ (m_{12}^4+\hat{s}^2)(3+\cos 2\theta)(4 \csc^2\Theta-2)\phantom{\sqrt{\hat{s}}} \right. \nonumber\\
&&  +8 m_{12}^2\,\hat{s}\, \sin^2\theta (2+\cos2\phi) \nonumber \\
&& \left.+8\,m_{12} \sqrt{\hat{s}}\,\left(m_{12}^2+\hat{s}\right)\cot\Theta\, \sin2\theta\, \cos\phi\right] \, ,\\
{\cal G}_2&=&16 \csc\Theta \left[(m_{12}^4+\hat{s}^2) \cos\theta \cot\Theta\right.  \nonumber\\
&& + \left. m_{12} \sqrt{\hat{s}}\left(m_{12}^2+\hat{s} \right)\ \sin\theta  \cos\phi \,\right]\, ,
\eea
where $g_{L(\ell)}$ and $g_{R(r)}$ are the $Z$ couplings to left- and right-handed quarks (leptons). 


In Fig.~\ref{fig:dist} we show the distributions in $\cos\Theta$, $\phi$, and $\cos\theta$ for  a 125~GeV Higgs boson and a background process  $d\bar{d}\to Z\gamma$ at $\sqrt{\hat{s}}=125$~GeV at the parton level. 
These are modified after including the effects of parton distribution functions ({PDF}) and detector acceptance and isolation cuts. 
In particular, we note that $\cos \Theta$ is directly connected to the photon $p_T$ through
%
\be
\cos\Theta = \sqrt{1-{4p_{\gamma T}^2\hat{s}}/{(\hat{s}-m_Z^2)^2}} \,.
\ee
The $\cos\Theta$ distribution in Fig.~\ref{fig:dist} therefore implies that the $p_{\gamma T}$ distribution is peaked at zero for the background and $(m_h^2-m_Z^2)/(2m_h)$ for the signal. However it also follows that once a cut on $p_{\gamma T}$ is imposed,
very little additional sensitivity can be gained from the $\cos\Theta$ distribution.

{\bf Analysis and Results} -- 
We perform  Monte Carlo simulations to obtain projections for the sensitivity
of this channel at the LHC using various analyses.
We consider Higgs masses of $120$, $125$, and $130$ GeV.  Our simulations
are specific to the $8$ and $14$ TeV LHC.  The existing $7$~TeV data has 
very little sensitivity in this channel such that we do not report those results here.

To perform these Monte Carlo studies, we generate at least 50,000
events for each signal and background process using MadGraph 5
~\cite{arXiv:1106.0522}. 
 The Higgs coupling to gluons and the $hZ\gamma$ vertex are implemented
as effective dimension five operators using the HEFT model provided by
MadGraph 5 and the FeynRules~\cite{Christensen:2008py} package. 
For both signal and background, the processes
$p p \to Z\gamma$ and $p p \to Z\gamma + 1j$ are generated, using the 
MLM matching scheme~\cite{MLM} implemented in MadGraph 5 and interfaced 
with Pythia~6~\cite{Pythiaref}, with a matching scale of $25$~GeV. 
Events are then passed to PGS 4~\cite{PGSref} using the CMS parameter card, 
to model detector acceptance and smearing effects. 

Since the energy and momentum resolution is crucial for this analysis, 
we have compared the invariant mass resolution obtained from PGS~4 with
the one that is obtained when smearing parton level events by hand using
the CMS detector parameters~\cite{Bayatian:2006zz}, and found 
that they agree in general. 


We demand that each lepton or photon has
\be
\label{eq:basicut}
|\eta| < 2.5 \quad {\rm and} \quad p_T > 15\  {\rm GeV}.
\ee  
The smearing results in the broadening of the lineshape in the total
invariant mass of the $Z\gamma$ system, $m_{\ell \ell \gamma}$, for the signal events.  Therefore,
before performing more detailed analyses, we perform an invariant mass
cut; demanding that the invariant mass of the $Z\gamma$ system be
within $5$ GeV of the mean invariant mass of the $Z\gamma$ system
in signal events. 
It is worth emphasizing that since subsequent analyses will effectively
reduce the range of invariant mass considered, the specific details
of this initial cut does not have a strong effect on the final value of
$S/\sqrt{B}$ obtained.
Note that this cut also effectively removes the background coming from FSR 
radiation that is characterized by $m_{\ell\ell\gamma} \sim M_Z$.

To determine the expected number of signal events at the $14$ TeV LHC, we obtain the 
inclusive Higgs production cross section from \cite{arXiv:1101.0593}.
For the $8$ TeV LHC, we use the values given 
in~\cite{Anastasiou:2012hx}.
The branching fraction for $h \to Z\gamma$ is found using 
HDECAY~\cite{hep-ph/9704448}, while we use the PDG value ($6.73\%$) for the 
branching fraction for a $Z$ decaying to  leptons
~\cite{FERMILAB-PUB-10-665-PPD}.  
The background cross section is found by using MCFM \cite{arXiv:1007.3492,Campbell:2011bn} with 
FSR photon radiation turned off.

\begin{figure}[t]
\includegraphics[scale=0.4, angle=0]{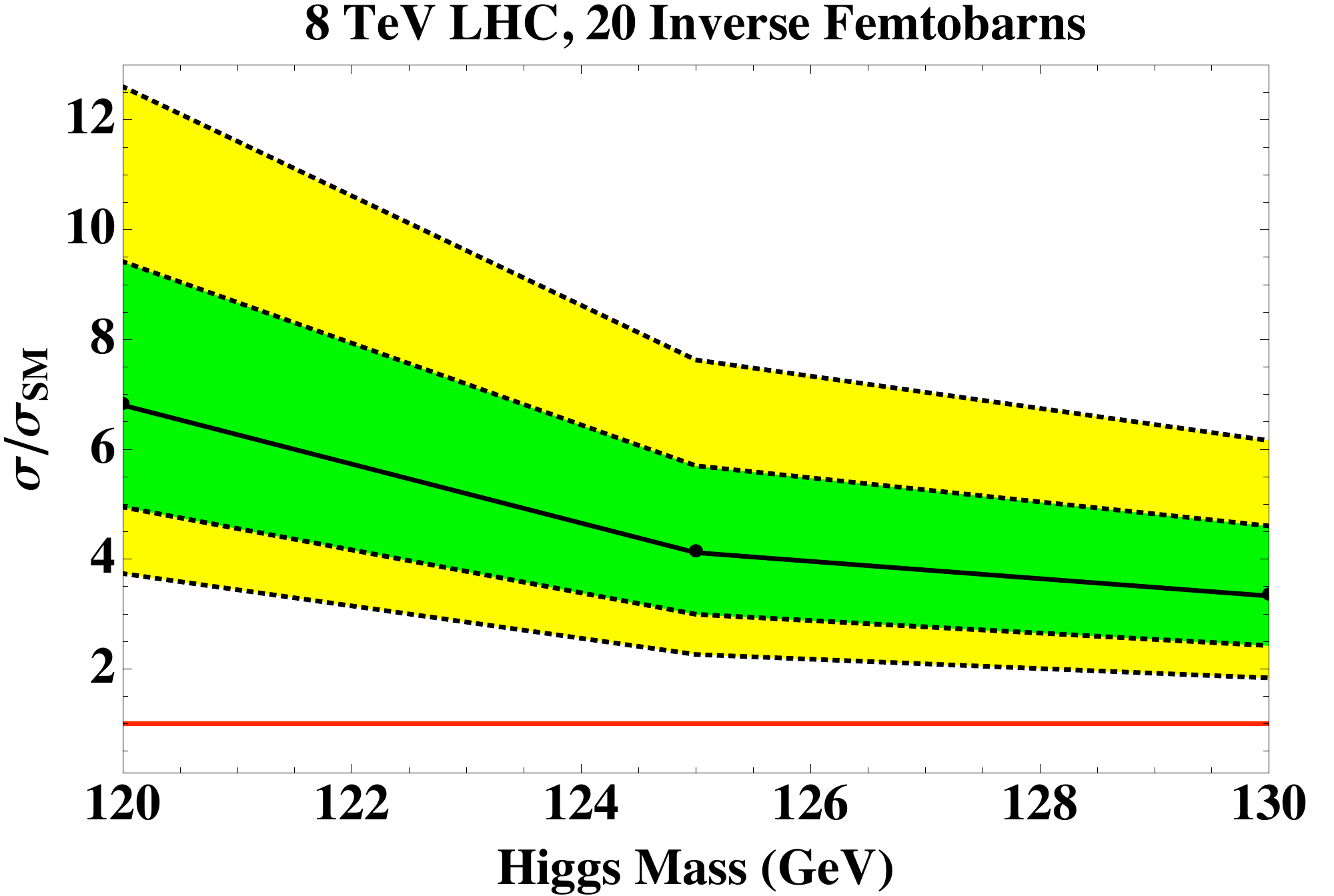}  
\caption{\label{fig2}{\em   Exclusion limits at the 95\% confidence level on the Higgs production rate times branching fraction to $Z\gamma$ at the
$8$ TeV LHC with an integrated luminosity of 20 fb$^{-1}$. The green (yellow) band is the 1(2) $\sigma$ contour.  The solid red line corresponds to the SM expectation.}}
\end{figure}

We perform three analyses, two of which are multivariate.  The multivariate discriminants we use are based on the matrix element of the signal and background processes. In the context of a maximum likelihood analysis such a discriminant was used in the discovery of the single top production in \cite{arXiv:0803.0739}. For simplicity we use a cut-based approach to determining our sensitivity using these multivariate discriminants.

We construct a discriminant using the fully differential cross sections computed for the signal and background processes to quantify the relative probability of a particular event being signal-like or background-like. We then determine an optimal cut on the discriminant to maximize the value for $S/\sqrt{B}$.   In one analysis, we include PDF weights for the leading initial state for signal or background events ($gg$ or $q\bar{q}$ respectively).  In the second multivariate analysis, we do not include 
a weight from PDFs. 
Labelling
the signal and background differential cross sections by $s(\mathbf{\Omega})$
and $b(\mathbf{\Omega})$, respectively, we consider the quantity
\begin{equation}
\label{eq:optcut}
D(\mathbf{\Omega})=\frac{s(\mathbf{\Omega})}{s(\mathbf{\Omega}) 
+ b(\mathbf{\Omega})} = 
\bigg( 1 + \frac{s(\mathbf{\Omega})}{b(\mathbf{\Omega})}\bigg)^{-1}.
\end{equation}
Here, $\mathbf{\Omega}=\{x_1, x_2, \hat{s}, m_{\ell\bar{\ell}}, \Theta, \theta, \phi\}$ is the complete set of kinematic  observables characterizing each event. When evaluating $D$ on a sample of pure signal events the distribution is peaked toward 1 while it is peaked toward $0$ for a pure background sample. For each Higgs mass, a cut on $D$ is determined by maximizing $S/\sqrt{B}$ of the events passing the cut. 
One advantage of using the multivariate discriminant in a cut-based approach is that the relative normalization of the signal and background cross sections does not affect the final significance computed using $S/\sqrt{B}$. The drawback, on the other hand, is that we lose those events not passing the cut, which would not be the case if the signal and background matrix elements were used to construct the likelihood directly.

Our multivariate discriminants use the parton-level differential cross section except for the Higgs propagator, 
as for the Higgs masses considered the Higgs width is much narrower than the experimental resolution.  In principle, one can deal with this issue by using transfer functions for the lepton momenta.  
We take the simpler approach
of weighting each event with a Gaussian invariant mass distribution that is centered at the average invariant mass for
signal events.  The width used in this Gaussian weighting is found
by scanning (in $20$ MeV increments) over potential values, from $100$ MeV to $5$ GeV, and selecting the value which maximizes the sensitivity of the analysis.   
The third analysis uses the same Gaussian invariant mass weight, but no other kinematic information about the events. While one would expect a loss of sensitivity, this approach has the advantage of being less sensitive to higher order corrections that could modify the angular distributions that enter the multivariate analyses. 



\begin{table}[t]
    \begin{tabular}{| l | r | r | r |}
      \hline
      Higgs Mass~ &  
      Signal (fb) & Backg. (fb) & $S/\sqrt{B}$ ($20$ fb$^{-1}$) \\ \hline
      $120$~GeV      &    $ 0.38 ~~ (0.45) $     &     $ 32. ~~ (110) $      &     $ 0.30 ~~ (0.19) $        \\ \hline
      $125$~GeV      &    $ 0.61 ~~ (0.74) $     &     $ 30. ~~ (100) $      &     $ 0.50 ~~ (0.33) $        \\ \hline
      $130$~GeV      &    $ 0.66 ~~ (0.86) $     &     $ 23. ~~ (89.) $      &      $ 0.62 ~~ (0.41) $        \\ \hline
   \end{tabular}
    \caption{
      \emph{
        The signal and background cross sections, as well as the significance after an optimal cut on the  discriminant in Eq.~(\ref{eq:optcut}) in the invariant mass only analysis at the $8$ TeV LHC. In the parenthesis we also show the corresponding values for all events passing the $p_{T}$ and geometric acceptance cuts and which are within an invariant mass window of $10$ GeV centered on the Higgs mass, as described in the text.}
    \label{table : analysis 8}
    }
\end{table}


\begin{table}[t]
 \center
    \begin{tabular}{| l | r | r | r |}
      \hline
      Higgs Mass~ &  
      Signal (fb) & Backg. (fb) & $S/\sqrt{B}$ ($100$ fb$^{-1}$) \\ \hline
      $120$~GeV      &    $ 0.83 ~~ (1.0) $     &       $ 36. ~~ (180) $      &     $ 1.2 ~~ (0.78) $        \\ \hline
      $125$~GeV      &    $ 1.3   ~~ (1.6) $     &       $ 37. ~~ (160) $      &     $ 2.0 ~~ (1.3) $        \\ \hline
      $130$~GeV      &    $ 1.7   ~~ (2.1) $     &       $ 40. ~~ (140) $      &     $ 2.7 ~~ (1.8) $        \\ \hline
   \end{tabular}
    \caption{
      \emph{Same as Tab.~\ref{table : analysis 8}, for the $14$~TeV LHC, with a luminosity of $100$~fb$^{-1}$.}
    }\label{table: analysis 14}
\end{table}

We find the best values for $S/\sqrt{B}$ from the analysis in which the full differential cross sections and PDF weights are used.  However the sensitivity from this analysis is only $\sim 1 \%$ larger than that obtained
from the invariant mass only analysis.  The smallness of this increase in sensitivity is due to the fact that the relatively hard $p_{\gamma T}$ cut leaves us without much additional sensitivity to $\Theta$, and the other angular variables are not as sensitive, especially given geometric acceptance and finite momentum resolution.  We therefore quote results using the invariant mass only analysis, as they should be more robust with respect to systematic uncertainties. In particular, the $m_{\ell\ell\gamma}$ distribution is unaffected by jet radiation, so that corrections to the jet multiplicity and momentum distribution, which is only simulated to leading order in our analysis, will not reduce the sensitivity. 

The signal and background cross sections after the optimal cut on $D$ from this invariant mass only analysis are listed in Table~\ref{table : analysis 8} for various  Higgs masses at the $8$ TeV LHC.  The expected significance with $20$ fb$^{-1}$ integrated luminosity is also provided.  
Table~\ref{table: analysis 14} shows analogous information for the $14$ TeV; here the expected significance with $100$ fb$^{-1}$ is shown.

\begin{figure}[t]
\includegraphics[scale=0.4, angle=0]{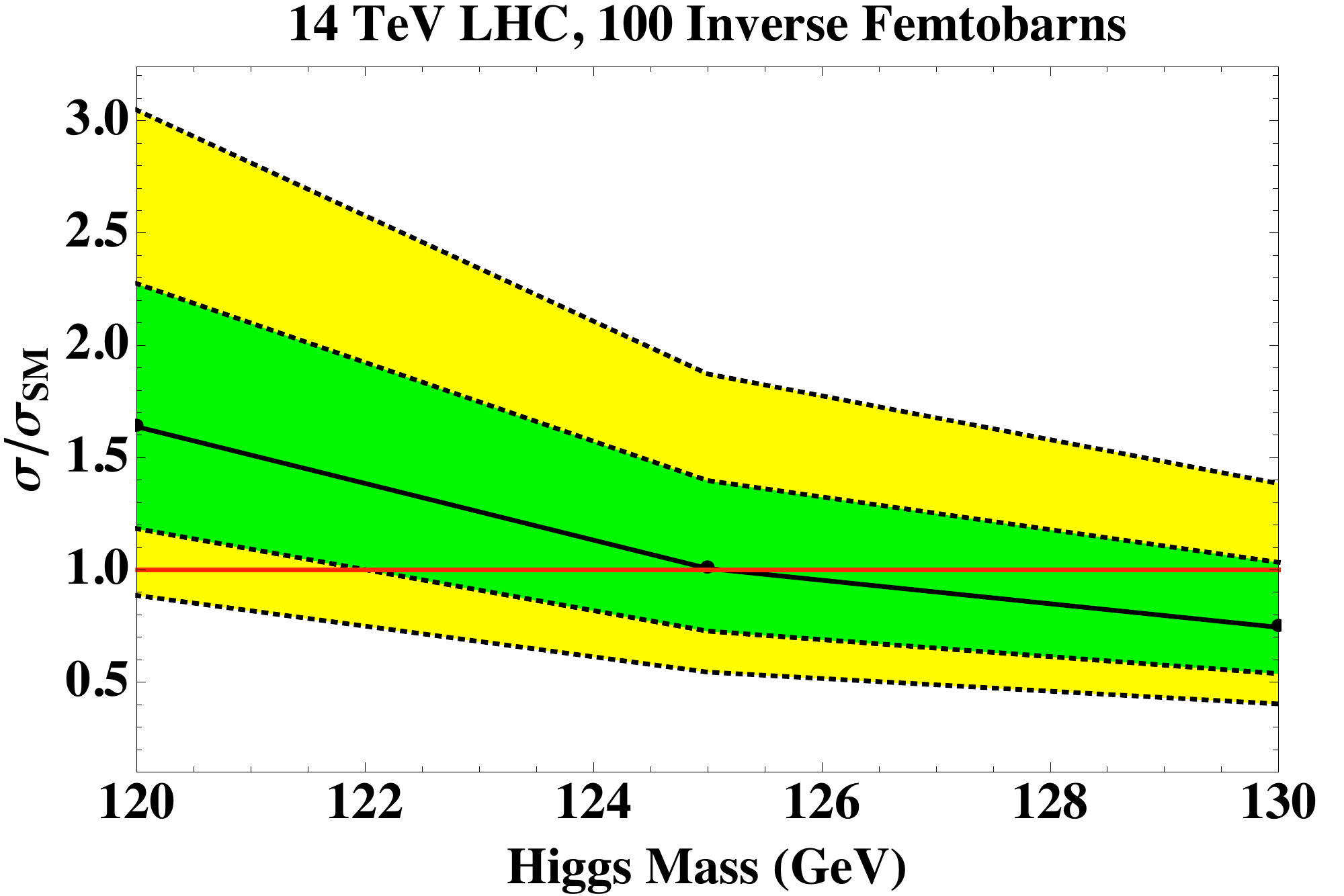}  
\caption{\label{fig3}{\em   Exclusion limits at the 95\% confidence level on the Higgs production rate times branching fraction to $Z\gamma$ at the
$14$ TeV LHC with an integrated luminosity of $100$ fb$^{-1}$. The green (yellow) band is the 1(2) $\sigma$ contour.  The solid red line corresponds to the SM expectation.}}
\end{figure}

In the absence of any signal, we have also considered the expected exclusion limit on the Higgs production rate in the gluon fusion channel using the CL$_s$ method \cite{CERN-OPEN-2000-205} with $20$ fb$^{-1}$ of integrated luminosity for the $8$~TeV LHC in Fig.~\ref{fig2} and for the $14$ TeV LHC with $100$ fb$^{-1}$ in Fig.~\ref{fig3}.


{\bf Conclusions} -- We have considered the possibility of searching for a light Higgs boson in its decays to $\ell\bar{\ell}\gamma$ final states via $Z\gamma$.  This branching ratio is known precisely in the SM, and deviations from this rate are unambiguous signals of new physics that couples to the Higgs boson, or could even signal the presence of a Higgs imposter~\cite{arXiv:1105.4587}. 

We have performed a detailed Monte Carlo study for the $8$ and $14$~TeV LHC. We find that branching ratios for the Higgs decay to $Z\gamma$ of several times the SM rate are probed at $8$ TeV with $20$ fb$^{-1}$, while the SM rate is probed at the $14$ TeV LHC with $100$ fb$^{-1}$. For Higgs masses of $125$~GeV and above, a measurement of the Higgs branching ratio to $Z\gamma$ is in reach of the $14$~TeV LHC. 
 We hope this work inspires experimental efforts in this particular search channel.
   

{\bf Acknowledgements} -- We benefitted from discussions with B. Auerbach, S. Chekanov, A. Kobach, H. Schellman, and M. Valesco.  The model file for Higgs to Z$\gamma$ decays was prepared by R. Vega-Morales; K. Kumar aided in the construction of our figures. I.L. and P.S. acknowledge the hospitality from the CERN TH-LPCC summer institute on LHC physics, where part  of this work was performed. This work was supported in part by the U.S. Department of Energy under
contract numbers DE-AC02-06CH11357,  DE-FG02-91ER40684, and DE-FG02-84ER40173.

\end{document}